\documentclass[12pt]{article}

\usepackage{epsfig}
\usepackage{amsfonts}
\usepackage{amsbsy}
\usepackage{amssymb}
\usepackage[mathscr]{eucal}
\def\be{\begin{equation}}\def\ee{\end{equation}}\def\vem{\vspace{1em}}
\def\De{\Delta} \def\tde{\tilde{\Delta}}\def\eps{\epsilon}\def\lal{\mathcal{L}}\def\al{\alpha}\def\dmu{\partial_\mu}\def\hence{\Rightarrow}\def\scrscr{\scriptscriptstyle}

\title{Giant gravitons as probes of gauged supergravity}
\author{Emily J. Hackett-Jones$^1$\footnote{email: e.j.hackett-jones@durham.ac.uk}
\quad and \quad
Douglas J. Smith$^2$\footnote{email: Douglas.Smith@durham.ac.uk} }

\begin{document}

\maketitle

\begin{center}

{\em Department of Mathematical Sciences,
University of Durham,
Science Laboratories,
South Road,
Durham. DH1 3LE.
UK}

\end{center}

\vspace{1.4cm}

\begin{abstract}
We consider giant graviton probes of 11-dimensional supergravity solutions
which are lifts of arbitrary solutions of 4-dimensional U(1)$^4$ and
7-dimensional U(1)$^2$ gauged supergravities. We show that the structure
of the lift ansatze is sufficient to explicitly find a solution for the
embedding and motion of the M5- or M2-brane in the internal space. The brane
probe action then reduces to that of a massive charged particle in the gauged
supergravity, demonstrating that probing the gauged supergravity with such
particles is equivalent to the more involved 11-dimensional brane probe
calculation. As an application of this we use these particles to probe
superstar geometries which are conjectured to be sourced by giant gravitons.
\end{abstract}

\vspace{-16cm}
\begin{flushright}
DCPT-02/59 \\
hep-th/0208015\\
\end{flushright}

\thispagestyle{empty}

\newpage

\setcounter{page}{1}

\section{Introduction}

We investigate giant gravitons in 4-dimensional U(1)$^4$ and
7-dimensional U(1)$^2$ gauged supergravities and their lifts to eleven
dimensions. From the point of view of the gauged supergravities, giant
gravitons are simply massive charged particles. However, from the
11-dimensional view-point, these particles can be described either as
massless particles or as branes wrapping (topologically trivial) spheres. These
extended objects are called giant gravitons \cite{McGreevy:2000cw}. The mass and charge of
such a particle in the lower dimensional gauged supergravity is given by the
angular momentum of the giant graviton in the internal space. The size of the
wrapped sphere grows with the angular momentum of the giant graviton. For
large angular momentum the giant graviton description can be trusted whereas
the massless particle would have a high concentration of energy which would
invalidate the probe approximation of neglecting the backreaction, and for
large enough angular momentum would even invalidate the low-energy supergravity
approximation of M-theory.

An interesting consequence of the relation between
the size of the sphere and angular momentum is that there is a maximum bound
on the angular momentum since the size of the sphere cannot exceed the size
of the internal space. Using the AdS/CFT correspondence this translates into
a bound on the R-charge of the class of operators dual to giant gravitons. It
turns out that this bound is given by $N$, the rank of the gauge group. Such
a bound is obvious in the field theory, e.g.\ for single-trace operators or
sub-determinant operators which are conjectured to be dual to giant gravitons
\cite{Balasubramanian:2001nh}. It
is not obvious that the gravity dual would encode such a finite-$N$ bound and
certainly the relation to a volume bound is novel. Note that this is also an
example of a UV/IR connection since a large angular momentum corresponds to
a large volume.

We study giant gravitons within the context of supergravity and in particular
the relation between gauged supergravities and 11-dimensional supergravity.
We show that
for any solution of the gauged supergravity, a probe calculation with a
massive charged particle is equivalent to probing the 11-dimensional lift
with an M5- or M2-brane. This extends the results of \cite{Page:2002xz} for the
closely related case of 5-dimensional U(1)$^3$ gauged supergravity.
Again it appears that the lift ansatze have precisely the right properties
for the existence of giant gravitons.

In \S~\ref{4d} we consider the case of 4-dimensional U(1)$^4$ gauged
supergravity. We first review the results of \cite{Cvetic:1999xp} for the
11-dimensional lift. In this case the giant graviton is an M5-brane
wrapping a 5-sphere in the internal space. It is supported from collapse
by coupling magnetically to the 4-form field strength $F_{(4)}$ of
11-dimensional supergravity. Before considering the M5-brane we show in
\S~\ref{ParticleProbe} that a massless 11-dimensional particle with
angular momentum is equivalent to the massive charged particles we wish to
consider in the gauged supergravity. We then move on to show that the same
massive charged particles have yet another description in terms of an M5-brane.
Specifically, in both cases we show that a massive charged particle probing any solution of
the gauged supergravity is equivalent to a massless particle or M5-brane
probing the 11-dimensional lift, with specific embeddings in the internal space.
In order to perform the M5-brane probe
calculation we first Hodge dualise $F_{(4)}$ in \S~\ref{StarF4} and then
integrate to find the 6-form potential, $*F_{(4)} = dA_{(6)}$, in
\S~\ref{IntF7}. In \S~\ref{probe4d} we then perform the giant graviton probe
calculation and show that the M5-brane action reduces to that of a massive
charged particle in four dimensions. This shows that probe calculations
performed with these particles in four dimensions are equivalent to first
lifting to eleven dimensions and then performing the M5-brane probe
calculations. Hence such calculations can be performed entirely within the
simpler lower dimensional context. As an application of this we probe superstar
backgrounds arising as the extremal limit of charged black holes
\cite{Duff:1999gh, Sabra:1999ux}, which are
conjectured to be sourced by these giant gravitons
\cite{Myers:2001aq, Balasubramanian:2001dx, Leblond:2001gn}.

In \S~\ref{7d} we repeat the above calculations for the 7-dimensional
U(1)$^2$ gauged supergravity and the corresponding superstar solutions
\cite{Cvetic:1999ne, Liu:1999ai}. We conclude with some comments in
\S~\ref{conclusions}.

\section{$D=11$ supergravity reduction on $S^7$}
\label{4d}
The compactification of 11-dimensional supergravity on $S^7$ leads to gauged $\mathcal{N}=8$ supergravity in four dimensions with gauge group SO(8). This theory arises from consistently truncating the massive Kaluza-Klein modes of the compactified 11-dimensional supergravity. Consequently all solutions of the 4-dimensional supergravity will correspond to solutions of the 11-dimensional theory. In practice, however, the relationship between solutions of the two theories is complicated and highly implicit. To provide a concrete realization of this relationship one can consistently truncate the 4-dimensional $\mathcal{N}=8$ theory to a $\mathcal{N} = 2$ theory. This corresponds to truncating the full gauge group SO(8) to its Cartan subgroup, $U(1)^4$. The explicit relationship between solutions of the $\mathcal{N} = 2$ theory and the 11-dimensional supergravity will be shown in the following.\vem

The $\mathcal{N} = 2$ supergravity theory has a bosonic sector consisting of the metric, four commuting U(1) gauge fields, three dilatons and three axions. We will be interested in cases where the axions are set to zero. While this is not completely consistent (since terms of the form $\eps_{\mu\nu\rho\sigma} F^{\mu\nu} F^{\rho\sigma}$ will source axions) it suffices for the present purposes since we will only consider electrically charged solutions. The Lagrangian for this theory is given by 
\be
e^{-1} \lal_4 = R - \frac{1}{2}(\partial \vec{\varphi})^2 + \frac{8}{L^2}(\cosh\varphi_1 + \cosh\varphi_2 + \cosh\varphi_3) - \frac{1}{4}\sum_{i=1}^4 e^{\vec{a_i}\cdot \vec{\varphi}}(F^i_{(2)})^2 \label{lag4}
\ee
Here $\vec{\varphi}=(\varphi_1, \varphi_2, \varphi_3)$ are the three dilaton fields, $F^i_{(2)} = dA_{(1)}^i$, $i=1,\ldots 4$, are the four U(1) field strength tensors, and the four 3-vectors $\vec{a_i}$ satisfy 
\be
M_{ij} = \vec{a_i}\cdot \vec{a_j} = 4\delta_{ij} -1 
\ee
The three dilaton fields can be conveniently parameterized in terms of four scalar quantities $X_i$, $i=1,\ldots 4$, where
\be
X_i = e^{-\frac{1}{2} \vec{a_i}\cdot \vec{\varphi}}
\ee
The $X_i$ satisfy the constraint $X_1 X_2 X_3 X_4 = 1$.
The Lagrangian (\ref{lag4}) leads to the equations of motion,
\begin{eqnarray}
d*_{(1,3)}d\log (X_i) &=& - \frac{1}{4}\sum_j M_{ij} X_j^{-2}*_{(1,3)}F^j_{(2)}\wedge F^j_{(2)} - \frac{1}{L^2}\sum_{jk}M_{ij}X_j X_k \eps_{(1,3)} \nonumber\\
&& + \frac{1}{L^2}\sum_j M_{ij} X_j^2\eps_{(1,3)} \\
d(X_i^{-2}*_{(1,3)}F^i_{(2)}) &=& 0 
\end{eqnarray}
together with the 4-dimensional Einstein-Maxwell equations coupled to the scalars $X_i$. 
Solutions of this 4-dimensional $\mathcal{N} = 2$ theory can be ``lifted'' to solutions of 11-dimensional supergravity as follows \cite{Cvetic:1999xp},
\be
ds_{11} = \tde^{2/3} ds_{(1,3)}^2 + \tde^{-1/3} \sum_i \left( L^2 X_i^{-1} d\mu_i^2 + X_i^{-1} \mu_i^2 (Ld\phi_i + A^i_{(1)})^2\right) \label{11metric}
\ee
where $\tde \equiv \sum_{i=1}^{4} X_i \mu_i^2$. The lift ansatz for the 4-form field strength tensor is
\be 
F_{(4)} = \frac{2U}{L}\eps_{(1,3)} - \frac{L}{2} \sum_i X_i^{-1} *_{(1,3)}dX_i \wedge d(\mu_i^2) + \frac{L}{2} \sum_i X_i^{-2} d(\mu_i^2) \wedge (L d\phi_i + A^i_{(1)})\wedge *_{(1,3)} F^i_{(2)} \label{4form}
\ee
where $U\equiv \sum_{i=1}^4 (X_i^2 \mu_i^2 - \tde X_i)$. Here $*_{(1,3)}$ means dualizing with respect to the metric $ds^2_{(1,3)}$. The four coordinates $\mu_i$ satisfy the constraint $\sum_i \mu_i^2 = 1$. They can be parameterized as
\be
\mu_1 = \cos\theta_1 \qquad \mu_2 = \sin\theta_1\cos\theta_2 \qquad \mu_3 = \sin\theta_1\sin\theta_2\cos\theta_3 \qquad \mu_4 = \sin\theta_1\sin\theta_2\sin\theta_3 \label{param}
\ee
where $0\leq \theta_i \leq \pi/2$ and $0\leq \phi_i \leq 2\pi$.

\subsection{A massless particle probe}
\label{ParticleProbe}

We consider a massless particle moving in the 11-dimensional space-time, Eq.~(\ref{11metric}), with some conserved angular momentum on the 7-sphere. It is interesting to consider how this particle appears in the associated 4-dimensional space-time. Clearly if the particle is stationary on the 7-sphere it will simply appear as a massless particle in four dimensions. However, if the particle has some angular momentum in the internal space we expect that it will behave as a massive charged particle in the 4-dimensional space-time. Here we show that this is indeed the case.\vem

For simplicity, we consider the action for a massive particle moving in eleven dimensions. We later take the mass to zero in the Hamiltonian formulation. The action is given by
\be
S = -m \int dt \sqrt{- \det\left(P(g)\right)}
\ee
where $P(g)$ is the pull-back of the 11-dimensional metric $ds_{11}^2$ onto the particle's world-line, i.e.
\be
P(g) = g_{AB}\dot{x}^A \dot{x}^B
\ee
where $x^A$, $A=0, \ldots 10$, are coordinates on the 11-dimensional space-time with $x^0 =t$ and $\dot{x}^A= dx^A/dt$. We assume that the motion on the 7-sphere is only in the $\phi_i$ directions, and that the particle is stationary in the $\mu_i$ directions. Therefore the Lagrangian is given explicitly by
\be
\lal = -m \left(-\tde^{2/3} g_{\mu\nu} \dot{x}^\mu \dot{x}^\nu - \tde^{-1/3} \sum_{i=1}^4 X_i^{-1} \mu_i^2 (L\dot{\phi}_i + A^i_\mu \dot{x}^\mu )^2\right)^{1/2}\ee
The momentum conjugate to $\phi_i$ can be easily computed for each $i=1, \ldots 4$. One obtains,
\be
P_i = \frac{\partial \lal}{\partial\dot{\phi}_i} = - \frac{m^2 L}{\tde^{1/3} \lal}\ X_i^{-1} \mu_i^2 (L \dot{\phi}_i + A^i_\mu \dot{x}^\mu)  \label{P_i}
\ee
Since the Lagrangian contains no explicit dependence on $\phi_i$ these momenta are time-independent. We want to rearrange Eq.~(\ref{P_i}) to write $\dot{\phi}_i$ in terms of the momenta, $P_j$. A few lines of algebra yields
\be
\dot{\phi}_i = \frac{\tde^{1/3} P_i X_i}{L^2 \mu_i^2} \left( \frac{-\tde\ g_{\mu\nu}\dot{x}^\mu\dot{x}^\nu}{\kappa + m^2 \tde^{1/3}} \right)^{1/2} - \frac{1}{L} A^i_\mu \dot{x}^\mu
\ee
where
\be
\kappa \equiv \sum_j \frac{\tde^{2/3}\ P_j^2 X_j}{L^2 \mu_j^2}
\ee
We now construct the Routhian,
\begin{eqnarray}
\mathcal{R} &=& \sum_i P_i \dot{\phi}_i - \lal\\
&=& \sum_i \frac{\tde^{1/3} P_i^2 X_i}{L^2 \mu_i^2} \left( \frac{-\tde\ g_{\mu\nu}\dot{x}^\mu\dot{x}^\nu}{\kappa + m^2 \tde^{1/3}} \right)^{1/2} - \frac{1}{L}\sum_i P_i A^i_\mu \dot{x}^\mu - \lal
\end{eqnarray}
In the limit $m \rightarrow 0$ the Routhian becomes
\be
\mathcal{R} = \sqrt{ -g_{\mu\nu} \dot{x}^\mu \dot{x}^\nu}\left( \sum_i \frac{\tde\ P_i^2 X_i}{L^2 \mu_i^2}\right)^{1/2} - \frac{1}{L} \sum_i P_i A^i_\mu \dot{x}^\mu  \label{mless}
\ee
We want to minimize the energy with respect to the coordinates $\mu_i$. This can be achieved by defining two 4-vectors $\mathbf{U}$ and $\mathbf{V}$ by
\be
U_i = \sqrt{\frac{X_i}{\mu_i^2} } \frac{P_i}{L} \qquad V_i = \sqrt{X_i \mu_i^2}
\ee
and recognizing that the quantity in brackets in Eq.~(\ref{mless}) can be written as
\be
\left(\sum_i \frac{\tde\ P_i^2 X_i}{L^2 \mu_i^2}\right)^{1/2} = |\mathbf{U}| |\mathbf{V}| \label{minthing}
\ee
By the Cauchy-Schwarz inequality the minimum value of this expression is $\mathbf{U}\cdot \mathbf{V} = \sum_i X_i P_i/L$ which occurs when $\mathbf{U}$ and $\mathbf{V}$ are parallel. The constraint $\sum_i \mu_i^2 =1$ determines the constant of proportionality relating $\mathbf{U}$ and $\mathbf{V}$ and we find that the minimal energy configuration occurs at $\mu_i^2 = P_i/ \sum_j P_j$. Thus, minimizing the energy in the compact directions produces the following Routhian
\be
\mathcal{R} = \frac{1}{L} \sum_i P_i X_i \sqrt{ -g_{\mu\nu} \dot{x}^\mu \dot{x}^\nu} - \frac{1}{L} \sum_i P_i A^i_\mu \dot{x}^\mu
\ee
This is just the Hamiltonian for a massive charged particle with scalar coupling moving in a 4-dimensional space-time with metric $ds_{(1,3)}^2 = g_{\mu\nu} dx^\mu dx^\nu$.

\subsection{Dualizing $F_{(4)}$}
\label{StarF4}

We are interested in probing the 11-dimensional solutions given in Eqs.~(\ref{11metric}) -- (\ref{4form}) with giant gravitons, which in this case are M5-branes with an $S^5$ topology. The 5-branes will couple to the 6-form potential, $A_{(6)}$, which is related to the 4-form field strength tensor $F_{(4)}$ via the dual field strength tensor, $F_{(7)}= * F_{(4)}= d A_{(6)}$. Therefore, to understand the motion of giant gravitons in the 11-dimensional background it is necessary to dualize the 4-form field strength tensor given in Eq.~(\ref{4form}) and then integrate it. Dualizing $F_{(4)}$ requires a number of results analogous to those obtained in the appendix of Ref.~\cite{Page:2002xz}, as described below. We will integrate $F_{(7)}$ in \S~\ref{IntF7}.\vem

First we consider dualizing a $(p+q)$-form of type $\al_{(p)} \wedge \beta_{(q)}$ in the 11-dimensional background, Eq.~(\ref{11metric}). Here $\al_{(p)}$ is a $p$-form in the ``AdS'' directions and $\beta_{(q)}$ is a $q$-form in the ``$S^7$'' directions. The following result will be useful,
\be
*_{(11)} (\al_{(p)} \wedge \beta_{(q)}) = (-)^{q(4-p)}\tde^{(1-4p+2q)/6}(*_{(1,3)}\al_{(p)} \wedge *_{(7)}\beta_{(q)}) \label{11simp}
\ee  
Here $*_{(7)}$ refers to the metric $ds_7^2$ where $ds_{11}^2 = \tde^{2/3} ds_{(1,3)}^2 + \tde^{-1/3}ds_7^2$, from Eq.~(\ref{11metric}). The metric on $S^7$ splits further into two parts; 
\be
ds_7^2 = L^2 \sum_i X_i^{-1} d\mu_i^2 + \sum_i X_i^{-1} \mu_i^2 (Ld\phi_i + A^i_{(1)})^2
\ee
Therefore, a result similar to Eq.~(\ref{11simp}) holds for dualizing forms in seven dimensions, namely
\be
*_{(7)} (\al_{(p)} \wedge \beta_{(q)}) = (-)^{q(3-p)} L^{3-2p} (*_{(3)} \al_{(p)} \wedge *_{(4)} \beta_{(q)})\label{7simp}
\ee  
where $\al_{(p)}$ is a $p$-form in the $\mu_i$ directions and $\beta_{(q)}$ is a $q$-form in the $\phi_i$ directions. Here $*_{(4)}$ refers to the metric $ds_4^2 = \sum_i X_i^{-1} \mu_i^2 (Ld\phi_i + A^i_{(1)})^2$ and $*_{(3)}$ refers to the metric 
\be
{d\tilde{s}}_4^2 =\sum_{i=1}^4 X_i^{-1} d\mu_i^2
\ee
restricted to the 3-sphere $S$: $\sum_{i=1}^4 \mu_i^2 = 1$. Due to this constraint on $\mu_i$, dualizing forms on $S$ is not completely straightforward and one needs the result
\be
*_{(3)} \al = \tilde{*}_{(4)}(e_4 \wedge \al) \label{43}
\ee
where $\tilde{*}_{(4)}$ refers to the metric ${d\tilde{s}}_4^2$ on $\mathbb{R}^4$ and $e_4 = \tde^{-1/2}\sum_i \mu_i d\mu_i$ is a unit 1-form normal to $S$. With the results Eqs.~(\ref{11simp})-(\ref{43}) we can now dualize $F_{(4)}$ in eleven dimensions.\vem

We define the following 2-forms on $S$, 
\begin{eqnarray}
Z_i &\equiv& \sum_{jkl} \eps_{ijkl}\ \mu_j d\mu_k \wedge d\mu_l\\
Z_{ij}&\equiv& \sum_{kl} \eps_{ijkl}\ d\mu_k\wedge d\mu_l
\end{eqnarray}
The volume form on $S$ is given by $W = \frac{1}{6}\eps_{ijkl}\ \mu_i d\mu_j\wedge d\mu_k \wedge d\mu_l$.
It can easily be shown that the 2-forms $Z_i$ and $Z_{ij}$  satisfy three identities:
\begin{eqnarray}
d Z_i &=& 6 \mu_i W \label{id1}\\
Z_i \wedge d\mu_j &=& -2(\delta_{ij} - \mu_i\mu_j)W \label{id2}\\
\sum_j X_j \mu_j Z_j \mu_i &=& \sum_j X_j \mu_j Z_{ji} + \tde Z_i \label{id3}
\end{eqnarray}
The 2-form $*_{(3)} d(\mu_i^2)$ can be written in terms of $Z_{ij}$ as follows,
\be
*_{(3)} d(\mu_i^2) = *_{(4)}\left(\tde^{-1/2} \sum_j \mu_j d\mu_j \wedge 2 \mu_i d\mu_i\right) = -\tde^{-1/2}\sum_j X_i X_j \mu_i \mu_j Z_{ij}
\ee
Similarly,
\begin{eqnarray}
*_{(3)}\ 1 = *_{(4)} \left(\tde^{-1/2} \sum_i \mu_i d\mu_i\right) &=& -\frac{\tde^{-1/2}}{6}\sum_{ij}X_j \mu_j d\mu_i \wedge Z_{ij} \nonumber\\
&=& -\frac{\tde^{1/2}}{6} \sum_i Z_i\wedge d\mu_i + \frac{\tde^{-1/2}}{6}\sum_{ij}\mu_i \mu_j X_j Z_j \wedge d\mu_i \nonumber\\
&=& \frac{\tde^{1/2}}{3} \sum_i (1- \mu_i^2)W - \frac{\tde^{-1/2}}{3} \sum_{ij}\mu_i \mu_j X_j (\delta_{ij} - \mu_i \mu_j)W \nonumber\\
\hence \qquad *_{(3)}\ 1&=& \tde^{1/2} W  \label{star1}
\end{eqnarray}
where we have used the identities Eqs.~(\ref{id3}) and (\ref{id2}) in the second and third steps respectively. Now we dualize the first term in $F_{(4)}$ using the results from Eqs~(\ref{11simp}), (\ref{7simp}) and (\ref{star1}),
\begin{eqnarray}
*_{(11)} \frac{2U}{L} \eps_{(1,3)} &=& - \frac{2U}{L}\ \tde^{-5/2}\ *_{(7)} 1\nonumber\\
&=& - 2U L^2\ \tde^{-5/2}\ *_{(3)} 1 \wedge *_{(4)} 1\nonumber\\
&=& -\frac{2 U L^2}{\tde^2}\ W \bigwedge_k \mu_k (L d\phi_k + A^k_{(1)}) 
\end{eqnarray}
Similarly, the results of Eqs.~(\ref{11simp})-(\ref{star1}) can be used to dualize the two other terms in $F_{(4)}$. One finds
\be
*_{(11)}\left( - \frac{L}{2} \sum_i X_i^{-1} *_{(1,3)} dX_i \wedge d(\mu_i^2)\right) = -\frac{L^2}{2\tde^2}\ \sum_{ij} X_j dX_i \wedge \mu_i \mu_j Z_{ij} \bigwedge_k \mu_k (Ld\phi_k + A^k_{(1)})
\ee
for the second term and 
\be
*_{(11)} \left( \frac{L}{2} \sum_i X_i^{-2} d(\mu_i^2) \wedge (L d\phi_i + A^i_{(1)}) \wedge *_{(1,3)} F^i_{(2)}\right) = \frac{L^2}{2\tde}\ \sum_{ij} F^i_{(2)} \wedge Z_{ij} X_j \mu_j \bigwedge_{k \neq i} \mu_k (Ld\phi_k + A^k_{(1)})
\ee
for the third term. Thus
\begin{eqnarray}
F_{(7)} = *_{(11)} F_{(4)} &=& -\frac{2L^2 U}{\tde^2} W \bigwedge_k \mu_k(L d\phi_k +A^k_{(1)}) - \frac{L^2}{2\tde^2}\sum_{ij} X_j dX_i\wedge \mu_i\mu_j Z_{ij} \bigwedge_{k} \mu_k(Ld\phi_k + A^k_{(1)}) \nonumber \\
&& +\frac{L^2}{2 \tde} \sum_{ij} F^i_{(2)} \wedge Z_{ij}X_j\mu_j \bigwedge_{k\neq i} \mu_k (Ld\phi_k +A^k_{(1)}) \label{F7}
\end{eqnarray}
It is now reasonably straightforward to check that $dF_{(7)} = 0$. However, one must take $F_{(2)}^i \wedge F_{(2)}^j = 0$ for this result to work. This corresponds to neglecting the axions, as already discussed. As expected, the result $dF_{(7)} = 0$ does not require use of the equations of motion because it is the Bianchi identity.

\subsection{Integrating $F_{(7)}$}
\label{IntF7}

We now wish to integrate $F_{(7)}$ obtained in Eq.~(\ref{F7}) to determine the 6-form potential $A_{(6)}$ which will couple to the giant graviton probes. Since $dF_{(7)}$ vanishes identically, such an $A_{(6)}$ must exist, at least locally. In fact we will find that it is not possible to determine $A_{(6)}$ globally, but it can be found locally.\vem

The first step is to rewrite the following 3-form using Eq.~(\ref{id3}),
\begin{eqnarray}
\tde^{-2}\sum_{ij} X_j dX_i\wedge \mu_i\mu_j Z_{ij} &=& \tde^{-1}\sum_i dX_i \wedge Z_i - \tde^{-2}\sum_{ij} X_j dX_i\wedge\mu_i\mu_j Z_{j}\mu_i\\
&=& \sum_i \dmu \left(\frac{X_i \mu_i}{\tde}\right) dx^\mu \wedge Z_i
\end{eqnarray}
Using this result the $2^{\rm nd}$ term in Eq.~(\ref{F7}) can be rewritten as follows,
\be
 \frac{L^2}{2\tde^2}\sum_{ij} X_j dX_i\wedge \mu_i\mu_j Z_{ij} \bigwedge_{l} \mu_l(Ld\phi_l + A^l) =  \frac{L^2}{2} \sum_i \dmu \left(\frac{X_i \mu_i}{\tde}\right) dx^\mu \wedge Z_i \bigwedge_l \mu_l (L d\phi_l + A^l) \label{rewriting}
\ee
Thus we postulate that the potential $A_{(6)}$ contains the following term 
\be
\tilde{A}_{(6)} = -\frac{L^2}{2} \sum_i \frac{X_i\mu_i}{\tde} Z_i \bigwedge_l \mu_l (L d\phi_l + A^l)
\ee
Evaluating $d\tilde{A}_{(6)}$ gives
\begin{eqnarray}
d\tilde{A}_{(6)} &=& -\frac{2L^2 U}{\tde^{2}} W\bigwedge_l \mu_l (L d\phi_l + A^l) - \frac{L^2}{2} \sum_i \dmu \left(\frac{X_i \mu_i}{\tde}\right)dx^{\mu} \wedge Z_i \bigwedge_l \mu_l (L d\phi_l + A^l) \nonumber \\
&& - \frac{L^2}{2} \sum_j Z_j \wedge F^j \bigwedge_{l\neq j} \mu_l (L d\phi_l + A^l) + \frac{L^2}{2\tde}\sum_{ij} F^i \wedge Z_{ij} X_j \mu_j\bigwedge_{l\neq i} \mu_l (L d\phi_l + A^l) \nonumber \\
&& - 6 L^2 W \bigwedge_l \mu_l(L d\phi_l + A^l) 
\end{eqnarray}
Therefore
\be
F_{(7)} = d\tilde{A}_{(6)} + \frac{L^2}{2} \sum_j Z_j \wedge F^j \bigwedge_{l\neq j}\mu_l (Ld\phi_l + A^l) +  6 L^2 W \bigwedge_l \mu_l (Ld\phi_l + A^l)
\ee
The sum of the last two terms in this expression is closed but not exact. For $\mu_1 \neq 0$ we can integrate them obtaining,
\be
d\left( \frac{L^2}{2} \frac{Z_1}{\mu_1} \bigwedge_j \mu_j (Ld\phi_j + A^j_{(1)}) + \frac{L^2}{2}\ \eps_{1jkl}\ \mu_k^2 \mu_l d\mu_l \wedge F^j_{(2)} \bigwedge_{ m \neq j} (L d\phi_m + A^m_{(1)})\right)
\ee
Then in the region where $\mu_1\neq 0$, $A_{(6)}$ is given by
\be
A_{(6)} = -\frac{L^2}{2\mu_1 \tde} \sum_i X_i \mu_i Z_{i1} \bigwedge_j \mu_j (L d\phi_j + A^j_{(1)}) + \frac{L^2}{2}\ \eps_{1jkl}\ \mu_k^2 \mu_l d\mu_l \wedge F^j_{(2)} \bigwedge_{ m \neq j} (L d\phi_m + A^m_{(1)}) \label{A6}
\ee
where we have used Eq.~(\ref{id3}) to replace the two terms involving $Z_i$ with one term involving $Z_{i1}$.

\subsection{Brane probe calculation}\label{probe4d}

We consider probing the 11-dimensional solution with giant gravitons. These giant gravitons are 5-branes which wrap an $S^5$ within the $S^7$. We choose the $S^5$ to be parameterized by the coordinates $\sigma^i = \{ \theta_2, \theta_3, \phi_2, \phi_3, \phi_4\}$. We consider the giant graviton moving rigidly in the $\phi_1$ direction at fixed $\theta_1$. The motion of the brane in the non-compact $AdS_4$ directions is arbitrary, but assumed to be independent of the brane world-volume coordinates, i.e.\ we consider rigid motion. The action for the giant graviton is,
\be
S_5 = - T_5 \int dt d\theta_2 d\theta_3 d\phi_2 d\phi_3 d\phi_4 \left[ \sqrt{- \det (P(g))} - \dot{x}^\mu A^{(6)}_{\mu \theta_2 \theta_3 \phi_2 \phi_3 \phi_4} - \dot{\phi}_1 A^{(6)}_{\phi_1 \theta_2 \theta_3 \phi_2 \phi_3 \phi_4} \right]
\ee
Here $P(g)$ is the pull-back of the 11-dimensional metric $ds_{11}^2$, Eq.~(\ref{11metric}), onto the world-volume of the 5-brane:
\be
P(g) = \gamma_{AB}  = g_{\al\beta} \frac{\partial X^{\al}}{\partial \sigma^A} \frac{\partial X^{\beta}}{\partial \sigma^B}
\ee
where $g_{\al\beta}$ is the 11-dimensional metric, $X^{\al}$ are the embedding coordinates of the brane in the 11-dimensional background and $\sigma^A = \{ t, \sigma^i\}$ are the world-volume coordinates of the brane. The 6-dimensional pull-back metric has non-zero entries along the diagonal and in the $(t, \phi_i)$ positions. Evaluating the determinant gives
\be
\det (\gamma_{AB}) = \frac{L^{10} X_1^2 \al}{\tde} \sin^{10} \theta_1 \cos^2 \theta_2 \sin^6\theta_2 \cos^2\theta_3\sin^2\theta_3 \left( g_{\mu\nu} \dot{x}^\mu \dot{x}^\nu + \frac{\cos^2\theta_1}{X_1\tde}\dot{\Phi}^2\right)
\ee
where $\al = X_2 \cos^2 \theta_2 + X_3 \sin^2\theta_2\cos^2\theta_3 + X_4 \sin^2\theta_2\sin^2\theta_3$.
The coupling of the 6-form potential to the probe world-volume can be determined simply by reading off the relevant components of $A_{(6)}$ from Eq.~(\ref{A6}). Using the parameterization for the $\mu_i$ given in Eq.~(\ref{param}) one finds 
\be
\dot{x}^\mu A^{(6)}_{\mu \theta_2 \theta_3 \phi_2 \phi_3 \phi_4} + \dot{\phi}_1 A^{(6)}_{\phi_1 \theta_2 \theta_3 \phi_2 \phi_3 \phi_4} = \frac{L^5}{\tde} \sin^6 \theta_1 \sin^3\theta_2 \cos\theta_2 \cos\theta_3\sin\theta_3\ \al\ \dot{\Phi} 
\ee 
where $\dot{\Phi} = L \dot{\phi}_1 + A^1_\mu \dot{x}^\mu$. Thus 
\begin{eqnarray}
S_5 &=& - T_5 L^5 \int dt d\theta_2 d\theta_3 d\phi_2 d\phi_3 d\phi_4  \frac{\sin^5\theta_1}{\sqrt{\tde}} \cos\theta_2 \sin^3\theta_2\cos\theta_3\sin\theta_3 \Bigg\{ - \frac{\sin\theta_1}{\sqrt{\tde}}\ \al\ \dot{\Phi} \nonumber \\ 
&& + \sqrt{X_1^2 \al\left( - g_{\mu\nu} \dot{x}^\mu \dot{x}^\nu - \frac{\cos^2\theta_1}{X_1\tde}\dot{\Phi}^2\right)}\ \Bigg\}
\end{eqnarray}
This action contains no explicit dependence on $\phi_1$. Thus we can replace all occurrences of $\dot{\phi}_1$ by the time-independent conjugate momentum $P_{\phi_{\scrscr 1}}(\theta_2, \theta_3, \phi_2, \phi_3, \phi_4)$. One obtains the following Routhian
\begin{eqnarray}
\mathcal{R} &=& \frac{1}{L} \sqrt{-g_{\mu\nu}\dot{x}^\mu\dot{x}^\nu}\Bigg(\frac{X_1 \tde}{\cos^2\theta_1}\left(P_{\phi_{\scrscr 1}}- \frac{N\al}{\tde} \sin^6\theta_1 \cos\theta_2\sin^3\theta_2\cos\theta_3\sin\theta_3\right)^2 \nonumber \\
&&\qquad + \frac{N^2 X_1^2\al}{\tde} \sin^{10}\theta_1 \cos^2\theta_2 \sin^6\theta_2\cos^2\theta_3\sin^2\theta_3\Bigg)^{1/2} -  \frac{P_{\phi_{\scrscr 1}} \dot{A}^1}{L} \label{Routhian}
\end{eqnarray}
where $N= T_5 L^6$. The terms inside the square root can be rewritten as a sum of squares:
\be
X_1^2 P_{\phi_{\scrscr 1}}^2 + X_1\al \tan^2\theta_1( P_{\phi_{\scrscr 1}} -  N\sin^4\theta_1\cos\theta_2\sin^3\theta_2\cos\theta_3\sin\theta_3)^2
\ee
This rearrangement makes it easy to minimize the energy over $\theta_1$. There are two minima occurring at $\theta_1 =0$ and $P_{\phi_{\scrscr 1}} = P_1 \cos\theta_2\sin^3\theta_2\cos\theta_3\sin\theta_3$, where $P_1$ is constant given by $ P_1= N \sin^4\theta_1$. The minimum at $\theta_1 =0$ corresponds classically to a massless particle, rather than a brane expanded on $S^5$. This solution is singular with respect to the gravitational field equations because it represents a huge amount of energy concentrated at a point, which leads to uncontrolled quantum corrections \cite{McGreevy:2000cw}. However, the latter minimum corresponds to a giant graviton. At this expanded minimum the Routhian reduces to 
\be
\mathcal{R} = \left(\frac{1}{L} \sqrt{-g_{\mu\nu}\dot{x}^\mu\dot{x}^\nu} X_1 P_1  - \frac{1}{L}P_1 A_\mu^1\dot{x}^\mu \right) \cos\theta_2\sin^3\theta_2\cos\theta_3\sin\theta_3
\ee
Integrating out the dependence on $\sigma^i = \{ \theta_2, \theta_3, \phi_2, \phi_3, \phi_4\}$ gives
\be
\mathcal{R} = \frac{1}{L} \sqrt{-g_{\mu\nu}\dot{x}^\mu\dot{x}^\nu}\ \tilde{P}_1 X_1  - \frac{1}{L}\tilde{P}_1 \dot{A}_\mu^1\dot{x}^\mu \label{result}
\ee
where $\tilde{P}_1 = P_1 V_5$ and $V_5 = \pi^3$ is the volume of a 5-dimensional sphere in flat space. This is just the Hamiltonian for a massive charged particle with scalar coupling moving in a 4-dimensional space-time with metric $ds_{(1,3)} =g_{\mu\nu}dx^\mu dx^\nu$. Equivalently we could have chosen the probe to move in any of the four $\phi_i$ directions. Then minimizing the energy over the remaining compact coordinate would give
\be
E_i = \frac{1}{L} \sqrt{-g_{\mu\nu}\dot{x}^\mu\dot{x}^\nu}\ \tilde{P}_i X_i  - \frac{1}{L} \tilde{P}_i A^i_\mu \dot{x}^\mu
\ee
for the energy of that probe.
So we find that by considering minimum energy configurations in the compact directions, the giant graviton action reduces to that of a charged particle coupled to a scalar field moving in four dimensions. This means that probing an 11-dimensional supergravity solution with giant gravitons is equivalent to probing the corresponding 4-dimensional solution with a charged particle. This result depends on the fact that the quantity under the square root in Eq.~(\ref{Routhian}) can be rearranged as a sum of squares. If this did not happen the minimization would be much more complicated and probably not produce such a simple result. 

\subsection{Probing superstars with giant gravitons}

In this section we use the giant graviton probe calculations of \S~\ref{probe4d} to probe a specific class of superstar geometries. Superstars are solutions of supergravity theories that arise by taking the supersymmetric limit of certain families of black hole solutions. In this limit the horizon disappears and the space-time is left with a naked singularity. It is thought that some superstar geometries may be sourced by giant gravitons, with the naked singularity interpreted physically as a collection of giant gravitons. Evidence for this was first given in Ref.~\cite{Myers:2001aq} where the authors considered type IIB superstar geometries and showed that the dipole field excited in the 5-form near the singularity corresponded to the dipole field excited by a collection of giant gravitons. Moreover, they argued that this ensemble of giant gravitons produced the correct mass and internal momentum for the superstar.\vem 

In this case we consider superstar solutions of 11-dimensional supergravity which are lifts of $\mathcal{N}=2$ supergravity in four dimensions with gauge group $U(1)^4$. The 4-dimensional theory admits the following 4-charge AdS black hole solution
\be
ds_4^2 = -\frac{f}{(H_1 H_2 H_3 H_4)^{1/2}} dt^2 + (H_1 H_2 H_3 H_4)^{1/2} (f^{-1} dr^2 + r^2 d\Omega_2^2)
\ee
where
\begin{eqnarray}
f &=& 1 - \frac{\mu}{r} + \frac{4 r^2}{L^2} H_1 H_2 H_3 H_4\\
H_i &=& 1 + \frac{q_i}{r}\\
A^i &=& (H_i^{-1} - 1)dt
\end{eqnarray}
We assume without loss of generality that the four U(1) charges $q_i$, $i=1, \ldots 4$, satisfy $q_1\geq q_2 \geq q_3 \geq q_4 \geq 0$. In the extremal limit, $\mu=0$, this solution has a naked singularity at $r=-q_4$. The apparent singularity in the metric at $r=0$ is a removable coordinate singularity. In the extremal limit we choose a new coordinate $\rho = r+q_4$ and extend the space-time past the coordinate singularity to $\rho =0$. This gives
\be
ds_4^2 = -\frac{\tilde{f}}{(\tilde{H}_1 \tilde{H}_2 \tilde{H}_3 \tilde{H}_4)^{1/2}} dt^2 + (\tilde{H}_1 \tilde{H}_2 \tilde{H}_3 \tilde{H}_4)^{1/2} (\tilde{f}^{-1} d\rho^2 + d\Omega_2^2)
\ee
where 
\begin{eqnarray}
\tilde{f} &=& (\rho - q_4)^2 + \frac{4}{L^2} \tilde{H}_1 \tilde{H}_2 \tilde{H}_3 \tilde{H}_4\\
\tilde{H}_i &=& \rho + q_i - q_4 \\
A^i &=& -\frac{q_i}{\tilde{H}_i}dt
\end{eqnarray}
In these coordinates the naked singularity occurs at $\rho =0$. The 4-dimensional solution can be lifted to an 11-dimensional supergravity solution via the lift ansatz given in Eqs.~(\ref{11metric}) and (\ref{4form}). The 11-dimensional solution will inherit a naked singularity from the 4-dimensional solution. We want to understand whether the naked singularity can be interpreted as a collection of giant gravitons. One way to test this idea is to probe the superstar geometry with giant gravitons. If the hypothesis is correct one expects that a probe carrying the same type of charge as the background will have a minimum energy configuration at the naked singularity, $\rho=0$. Moreover, we expect $\rho=0$ to be a BPS minimum ($E_i = \frac{P_i}{L}$) for this type of probe. We consider giant graviton probes carrying momentum in the $\phi_i$ direction and wrapping the $\theta_{j\neq i}$, $\phi_{j\neq i}$ directions. We look for solutions which are stationary in the extended directions, {\it i.e.} $\dot{x}^\mu =0$ except $\mu=0$. From Eq.~(\ref{result}) the energy of such a probe is given by
\be
E_i = \frac{1}{L}\sqrt{-g_{00}}\ \tilde{P}_i X_i - \frac{\tilde{P}_i A^i}{L} = \frac{\tilde{P}_i}{L} \frac{\tilde{f}^{1/2} + q_i}{\tilde{H}_i}
\ee
There are five distinct cases for the background charge, which we consider in turn:  
\begin{enumerate}
\item all $q_i =0$, then BPS minimum for all 4 types of probe at $\rho=0$.
\item $q_1 \neq 0$, all other $q_i =0$. The probe coupling to $A^1$ has a BPS minimum at $\rho =0$. Energy of probes coupling to $A^2$, $A^3$, $A^4$ saturates the BPS bound at $\rho =0$, but the gradient of the potential is non-zero at $\rho =0$.
\item $q_1, q_2 \neq 0$, all other $q_i =0$. Energy of probes coupling to $A^1$, $A^2$ saturates the BPS bound at $\rho =0$, but the gradient of the potential is non-zero. Probes coupling to $A^3$, $A^4$ neither saturate the BPS bound, nor have a minimum at $\rho=0$. 
\item $q_1, q_2, q_3 \neq 0$, $q_4 =0$. Energy of probes coupling to $A^1$, $A^2$, $A^3$ saturates the BPS bound at $\rho=0$, but gradient of potential is infinite. Energy of probe coupling to $A^4$ diverges as $\rho\rightarrow 0$.
\item $q_1, q_2, q_3, q_4 \neq 0$. Energy of probe coupling to $A^4$ diverges as $\rho\rightarrow 0$. The gradient of the potential for probes coupling to $A^1$, $A^2$, $A^3$ is non-zero at $\rho=0$. 
\end{enumerate}
Thus we find that the only scenario where it is sensible to interpret the naked singularity as a configuration of giant gravitons is in the singly charged background. Here the probe which carries the same type of charge as the background has a BPS minimum at $\rho=0$. In all other cases the energy of the probe is not minimized at the singularity. This is similar to what was found in Ref.~\cite{Page:2002xz} in the context of type IIB supergravity, where only the singly charged background could be interpreted in terms of giant gravitons.

%\newpage
\section{Supergravity reduction on $S^4$}
\label{7d}

The Kaluza-Klein reduction of 11-dimensional supergravity on $S^4$ leads to $\mathcal{N} = 4$ supergravity in seven dimensions with gauge group SO(5). As in the previous case, this $\mathcal{N} = 4$ theory can be consistently truncated to $\mathcal{N} = 2$ supergravity coupled to a vector multiplet. The vector multiplet consists of the metric, a 2-form potential, four vector potentials and four scalars. We are interested in a further truncation of the 7-dimensional theory where only the metric, two vector potentials and two scalars are retained in the bosonic sector. That is, the only gauge fields retained are those corresponding to the $U(1)^2$ Cartan subgroup of SO(5). Like the previous case, where we neglected axions, this further truncation is not completely consistent. However, solutions which preserve $F^1_{(2)}\wedge F^2_{(2)} =0$ will be solutions of both the full and truncated $\mathcal{N} = 2$ theories. These solutions will also correspond directly to solutions of 11-dimensional supergravity. The lift ansatz will be described in the following paragraph.\vem

The Lagrangian for the 7-dimensional truncated $\mathcal{N} = 2$ theory is given by
\begin{eqnarray}
e^{-1}\mathcal{L}_7&=& R - \frac{1}{2}\sum_{i=1}^2 (X_i^{-1}\partial X_i)^2 - \frac{1}{4}(X_0^{-1}\partial X_0)^2 + \frac{1}{L^2}(4X_1 X_2 + 2X_0 X_1 + 2X_0 X_2 - \frac{1}{2}X_0^2)\nonumber\\
&& - \frac{1}{4}\sum_{i=1}^2 X_i^{-2}(F^i_{(2)})^2
\end{eqnarray}
Here $X_0, X_1, X_2$ are a convenient parameterization for the two scalar fields. They obey the constraint $X_0 = (X_1 X_2)^{-2}$. The 2-forms $F^i_{(2)} = dA^i_{(1)}$, $i=1,2$ are field strengths for the U(1) gauge groups. The 7-dimensional equations of motion can be easily deduced from the above Lagrangian. We obtain \footnote{From now on we will use $\{ \al,\beta, \gamma, \ldots\}$ for indices running over 0, 1, 2, while $\{i, j, k, \ldots \}$ run over 1, 2 only.}
\begin{eqnarray}
d*_{(1,6)} d \log(X_i) &=& - \frac{4 X_0^{-1/2}}{L}\ \eps_{(1,6)} - \frac{2X_0 X_i }{L^2}\eps_{(1,6)} - X_i^{-2} F^i \wedge *_{(1,6)} F^i - 2\lambda\\
d*_{(1,6)} d \log(X_0) &=& - \frac{4 X_0}{L^2} \sum_i X_i \eps_{(1,6)} + \frac{2 X_0^2}{L^2} \eps_{(1,6)} - 2 \lambda\\
d \left(X_i^{-2} *_{(1,6)} F^i\right) &=& 0
\end{eqnarray}
where
\be
\lambda = \frac{1}{5L^2} \left( -8 X_0^{-1/2} - 4X_0\sum_i X_i + X_0^2\right)\eps_{(1,6)} - \frac{1}{5} \sum_i X_i^{-2} F^i\wedge *_{(1,6)} F^i
\ee
and  $*_{(1,6)}$ means dualizing with respect to the metric $ds_{(1,6)}^2$.
Solutions of the equations of motion can be lifted to solutions of 11-dimensional supergravity via the lift ansatz:
\begin{eqnarray}
ds_{11}^2 &=& \De^{1/3} ds_{(1,6)}^2 + \De^{-2/3}\left( L^2 \sum_{\al=0}^2 X_\al^{-1}d\mu_\al^2 + \sum_{i=1}^2 X_i^{-1} \mu_i^2(Ld\phi_i +A^i)^2\right) \label{11metric2}\\
F_{(7)} &=& - \frac{2U}{L}\eps_{(1,6)} - \frac{1}{L}\De X_0 \eps_{(1,6)} - \frac{L}{2}\sum_{\al =0}^2 X_\al^{-1}*_{(1,6)}dX_\al \wedge d(\mu_\al^2) \nonumber\\
&& \qquad - \frac{L}{2}\sum_{i=1}^2 X_i^{-2} d(\mu_i^2)\wedge (Ld\phi_i + A^i_{(1)}) \wedge *_{(1,6)}F^i_{(2)} \label{7form}
\end{eqnarray}
where $U = \sum_{\al=0}^2 (X_\al^2 \mu_\al^2 - \De X_\al)$ and $\De = \sum_{\al =0}^2 X_\al \mu_\al^2$. The variables $\mu_\al$ satisfy the constraint $\sum_\al \mu_\al^2 =1$. They can be parameterized by 
\be
\mu_0 = \sin\theta_1\cos\theta_2, \qquad \mu_1= \cos\theta_1, \qquad \mu_2 = \sin\theta_1\sin\theta_2, \label{param2}
\ee
where $0\leq \theta_1, \theta_2 \leq \pi/2$ and $0\leq \phi_1, \phi_2 \leq 2\pi$.

\subsection{Dualizing $F_{(7)}$}
We want to consider probing the 11-dimensional supergravity solutions Eqs.~(\ref{11metric2})-(\ref{7form}) with giant gravitons. As in \S~\ref{ParticleProbe} we could also consider probing with a massless particle moving on the 4-sphere. We find that this is equivalent to probing the associated 7-dimensional space-time with a massive charged particle. However, we do not repeat the calculation here. The giant gravitons used here to probe the solution will be M2-branes with an $S^2$ topology. The 2-branes will couple to a 3-form potential $A_{(3)}$, related to the 7-form field strength tensor $F_{(7)}$ via the dual field strength $F_{(4)}= *_{(11)} F_{(7)}= dA_{(3)}$. Therefore to find $A_{(3)}$ explicitly we need to dualize $F_{(7)}$, given in Eq.~(\ref{7form}), and integrate the resulting 4-form. In many ways this is similar to the previous case where a 4-form field strength tensor was dualized (\S~\ref{StarF4}) and then integrated (\S~\ref{IntF7}) to find a 6-form potential $A^{(6)}$. The main differences arise in the intermediate calculations because here the sphere is even-dimensional, and thus parameterized slightly differently compared to $S^7$. However, the end results will be largely similar.\vem

In analogy with the previous case, we first need a result for dualizing $(p+q)$-forms which split into a product of a $p$-form, $\al_{(p)}$, in the ``AdS'' directions and a $q$-form, $\beta_{(q)}$, in the ``$S^4$'' directions. We find
\be
*_{(11)} (\al_{(p)}\wedge \beta_{(q)}) = (-)^{q(7-p)}\De^{(-2p+4q -1)/6}(*_{(1,6)}\al_{(p)} \wedge *_{(4)} \beta_{(q)}) \label{res1}
\ee
The metric on $S^4$ also splits into two parts: 
\be
ds_4^2 =  L^2 \sum_{\al=0}^2 X_\al^{-1}d\mu_\al^2 + \sum_{i=1}^2 X_i^{-1} \mu_i^2(Ld\phi_i +A^i)^2
\ee
Therefore, we have a result similar to Eq.~(\ref{res1}) for dualizing forms within the $S^4$,
\be
*_{(4)}(\al_{(p)}\wedge \beta_{(q)}) = (-)^{q(2-p)+1}L^{2-2p}(\tilde{*}_{(2)}\al_{(p)}\wedge *_{(2)}\beta_{(q)}) \label{res2}
\ee
where $\al_{(p)}$ is a $p$-form in the $\mu_\al$ directions and $\beta_{(q)}$ is a $q$-form in the $\phi_i$ directions. Here $*_{(2)}$ means dualizing with respect to $ds_2^2 = \sum_i X_i^{-1} \mu_i^2(Ld\phi_i +A^i)^2$, whereas $\tilde{*}_{(2)}$ refers to the metric 
\be
ds_3^2 = \sum_{\al=0}^2 X_\al^{-1} d\mu_\al^2 \label{ds3}
\ee
restricted to the surface $S: \sum_\al \mu_\al^2 =1$. Dualizing forms on $S$ requires the following result \cite{Page:2002xz},
\be
\tilde{*}_{(2)} \al = *_{(3)} (e_3 \wedge \al)
\ee
where $*_{(3)}$ refers to $ds_3^2$ in Eq.~(\ref{ds3}) and $e_3=\De^{-1/2} \sum_\al \mu_\al d\mu_\al$ is a unit 1-form orthogonal to $S$. We now have all the necessary results to dualize $F_{(7)}$ in eleven dimensions.\vem
  
It is useful to define the following 1-forms \cite{Page:2002xz},
\begin{eqnarray}
Z_{\al\beta} &=& \eps_{\al\beta\gamma}d\mu_\gamma\\
Z_{\al} &=& \eps_{\al\beta\gamma}\mu_\beta d\mu_\gamma
\end{eqnarray}
Due to the constraint $\sum_\al \mu_\al^2=1$, there are three identities connecting these 1-forms \cite{Page:2002xz}:
\begin{eqnarray}
dZ_\al &=& 2 \mu_\al W \label{ident1}\\
Z_\al \wedge d\mu_\beta &=& (\delta_{\al\beta} - \mu_\al \mu_\beta)W\\
\sum_\al X_\al \mu_\al Z_{\al\beta} &=& \sum_\al X_\al \mu_\al Z_\al \mu_\beta - \De Z_\beta \label{ident3}
\end{eqnarray}
where $W =\frac{1}{2}\eps_{\al\beta\gamma}\mu_\al d\mu_\beta\wedge d\mu_\gamma$ is the volume form on $S$. Using these identities we can obtain some intermediate results:
\begin{eqnarray}
\tilde{*}_{(2)} d\mu_\al &=& \tilde{*}_{(3)} (\De^{-1/2}\sum_{\beta}\mu_\beta d\mu_\beta\wedge d\mu_\al) = - X_0^{-1/4} \De^{-1/2}\sum_\beta X_\al Z_{\al\beta}\mu_\beta X_\beta \\
\tilde{*}_{(2)} 1 &=& \tilde{*}_{(3)} (\De^{-1/2}\sum_{\al}\mu_\al d\mu_\al) = X_0^{-1/4} \De^{1/2} W \label{last}
\end{eqnarray}
The factors of $X_0$ in these expressions arise from the relation $X_0 = (X_1 X_2)^{-2}$, which means that the determinant of the metric $ds_3^2$, given in Eq.~(\ref{ds3}), is $X_0^{-1/2}$ (c.f. previous case, where $\det(d\tilde{s}_4^2) = (X_1 X_2 X_3 X_4)^{-1} =1$). However, these extra factors of $X_0$ will cancel out when we dualize terms in $F_{(7)}$. For example, using Eqs.~(\ref{res1}), (\ref{res2}) and (\ref{last}), we can dualize the first term in $F_{(7)}$ as follows,
\begin{eqnarray}
*_{(11)} \left( \frac{2U}{L} \eps_{(1,6)}\right) &=& -\frac{2U}{L} \De^{-5/2} *_{(4)} 1 \nonumber \\
&=& 2U L  \De^{-5/2} \tilde{*}_{(2)} 1 \wedge *_{(2)} 1 \nonumber \\
&=& 2U L  \De^{-2} X_0^{-1/4} W \bigwedge_i X_i^{-1/2}\mu_i(Ld\phi_i + A^i_{(1)}) \nonumber\\
&=& 2U L  \De^{-2} W \bigwedge_i \mu_i(Ld\phi_i + A^i_{(1)})
\end{eqnarray}
where we have used the constraint $X_0 = (X_1 X_2)^{-2}$ in the last line. Similarly we can dualize the other terms in $F_{(7)}$ using the results Eqs.~(\ref{res1})-(\ref{last}). We find
\begin{eqnarray}
F_{(4)} = *_{(11)} F_{(7)} &=& \frac{2LU}{\De^2} W\bigwedge_i \mu_i(Ld\phi_i + A^i_{(1)}) + \frac{L X_0 }{\De} W\bigwedge_i \mu_i (Ld\phi_i + A^i_{(1)}) \nonumber \\
&&+\frac{L}{\De^2}\sum_{\al, \beta} \mu_\al dX_\al \wedge Z_{\al\beta}\mu_\beta X_\beta\bigwedge_i \mu_i(Ld\phi_i + A^i_{(1)}) \nonumber \\
&& + \frac{L}{\De}\sum_{i\beta} F^i_{(2)}\wedge Z_{i\beta}\mu_\beta X_\beta \bigwedge_{j\neq i} \mu_j (Ld\phi_j + A^j_{(1)}) \label{F4}
\end{eqnarray}
It is straightforward to show that $d F_{(4)} =0$, using the identities given in Eqs.~(\ref{ident1})-(\ref{ident3}). This means that $F_{(4)}$ can be integrated at least locally.

\subsection{Integrating $F_{(4)}$}
The procedure for integrating $F_{(4)}$ is very similar to the previous case. Therefore, we will not repeat the calculation here, but just show the final result.  As before, it is not possible to write $F_{(4)} = dA_{(3)}$ with $A_{(3)}$ well-defined over the whole space-time. However, $A_{(3)}$ can be found locally everywhere. For example, in the region where $\mu_1 \neq 0$, $A_{(3)}$ is given by
\be
A_{(3)} =  \frac{L}{\mu_1 \De}\ \sum_\al \mu_\al X_\al Z_{\al 1}\bigwedge_i \mu_i (Ld\phi_i + A^i_{(1)}) + L \mu_0 F^2_{(2)} \wedge (Ld\phi_1 + A^1_{(1)}) \label{A3}
\ee
In the next section we will consider giant graviton probes moving in the 11-dimensional space-time at fixed $\mu_1 \neq 0$. The above form for $A_{(3)}$ will allow the coupling of the probe to the 3-form potential to be determined explicitly.

\subsection{Brane probe calculation}
We consider probing the 11-dimensional solution with a giant graviton. In this case the giant graviton is a 2-brane wrapped on an $S^2$ within the $S^4$. We take the brane world-volume to be parameterized by the coordinates $t, \theta_2, \phi_2$ and consider rigid motion in the $\phi_1$ direction at fixed $\theta_1$. The motion in the non-compact AdS directions is arbitrary, but assumed to be independent of the coordinates $\theta_2$, $\phi_2$, so that only rigid motion of the brane is considered. The action for the giant graviton is given by
\be
S_2 = -T_2\int dt d\theta_2 d\phi_2\left[ \sqrt{-\det(P(g))} + \dot{x}^\mu A^{(3)}_{\mu\theta_2\phi_2} + \dot{\phi}_1 A^{(3)}_{\phi_1 \theta_2\phi_2} \right]
\ee
As before, $P(g)$ is the pull-back of the 11-dimensional metric onto the 3-dimensional world-volume of the brane. The determinant of this metric can be evaluated readily. One obtains,
\be
\det(P(g)) =  \De^{-1} L^4 X_1^2 \sin^4\theta_1 \sin^2\theta_2 \al\left(g_{\mu\nu} \dot{x}^\mu \dot{x}^\nu + \frac{\cos^2\theta_1}{X_1\De}\dot{\Phi}^2\right)
\ee
The components of $A_{(3)}$ which couple to the brane can be read off from Eq.~(\ref{A3}), using the parameterization for $\mu_\al$ given in Eq.~(\ref{param2}). We obtain,
\be
\dot{x}^\mu A^{(3)}_{\mu\theta_2\phi_2} + \dot{\phi}_1 A^{(3)}_{\phi_1 \theta_2\phi_2} = - \frac{L^2}{\De} \sin^3\theta_1 \sin\theta_2 \al \dot{\Phi} 
\ee
where $\al = X_0 \cos^2\theta_2 + X_2\sin^2 \theta_2$ and $\dot{\Phi} = L\dot{\phi_1} + A^{1}_{\mu}\dot{x}^\mu$. Thus we get the following Lagrangian for the giant graviton,
\be
\lal = -T_2 L^2\left\{ \frac{X_1}{\sqrt{\De}} \sin^2\theta_1\sin\theta_2\al^{1/2} \sqrt{-g_{\mu\nu} \dot{x}^\mu \dot{x}^\nu - \frac{\cos^2\theta_1}{X_1\De}\dot{\Phi}^2} - \frac{1}{\De}\sin^3\theta_1\sin\theta_2\al\dot{\Phi}\right\}
\ee
As in the previous case, there is no explicit dependence on $\phi_1$ in the Lagrangian and so we replace $\dot{\phi_1}$ with its time-independent conjugate momentum $P_{\phi_1}(\theta_2, \phi_2)$. This yields the following Routhian,
\begin{eqnarray}
\mathcal{R} &=& \frac{1}{L}\sqrt{-g_{\mu\nu} \dot{x}^\mu \dot{x}^\nu}\left(\frac{X_1\De}{\cos^2\theta_1}\left(P_{\phi_1} - \frac{N}{\De}\sin^3\theta_1\sin\theta_2\al\right)^2 + \frac{N^2 X_1^2\al}{\De} \sin^4\theta_1\sin^2\theta_2\right)^{1/2} \nonumber \\
&& - \frac{1}{L} P_{\phi_1}A^{1}_\mu \dot{x}^\mu \label{Routh2}
\end{eqnarray}
As before, the quantity in the square root can be rearranged as a sum of squares to give
\be
\mathcal{R} = \frac{1}{L}\sqrt{-g_{\mu\nu} \dot{x}^\mu \dot{x}^\nu}\left( X_1^2 P_{\phi_1}^2 + X_1\al \tan^2\theta_1(P_{\phi_1} - N\sin\theta_1\sin\theta_2)^2\right)^{1/2} - \frac{1}{L} P_{\phi_1}A^{1}_\mu \dot{x}^\mu
\ee
It is now simple to minimize the energy over $\theta_1$. There are two minima:
$\theta_1 =0$ and $P_{\phi_1} = P_1 \sin\theta_2$, where $P_1 = N\sin\theta_1$ is constant. Like the previous case, the minimum at $\theta_1 =0$ is singular as it represents a huge energy concentrated at a point. From now on we consider the latter minimum which represents a giant graviton. At this minimum the Routhian becomes
\be
\mathcal{R} = \frac{1}{L}\sqrt{-g_{\mu\nu} \dot{x}^\mu \dot{x}^\nu}\ X_1 \tilde{P}_1 - \frac{1}{L} \tilde{P}_1 A^{1}_\mu \dot{x}^\mu
\ee
after integrating out $\theta_2, \phi_2$ and setting $\tilde{P}_1 = 2\pi  P_1$. This is the Hamiltonian for a massive charged particle coupled to a scalar in seven dimensions.
Equivalently we could have chosen the brane to wrap the $\theta_1, \phi_1$ directions and move in the $\phi_2$ direction. This would produce an entirely analogous result. Thus in general the energy of a probe moving in the $\phi_i$ direction is given by
\be
E_i = \frac{1}{L}\sqrt{-g_{\mu\nu} \dot{x}^\mu \dot{x}^\nu}\ X_i \tilde{P}_i - \frac{1}{L} \tilde{P}_i A^{i}_\mu \dot{x}^\mu \label{Ei2}
\ee 
where $i=1,2$. Therefore we have shown that probing the 11-dimensional solutions Eqs.~(\ref{11metric2})-(\ref{7form}) with a giant graviton is equivalent to probing the related 7-dimensional geometry with a massive charged particle. As before, this result depends on the fact that the quantity in the square root in Eq.~(\ref{Routh2}) can be rearranged as a sum of squares to simplify the minimization procedure.

\subsection{Probing superstars with giant gravitons}
We now consider probing particular superstar geometries with giant gravitons. These superstars are lifts of $\mathcal{N} =2$ supergravity in seven dimensions with gauge group $U(1)^2$. The 7-dimensional theory admits the following family of black hole solutions with two charges,
\be
ds_7^2 = - \frac{f\ r^{2/5}}{(H_1 H_2)^{4/5}} dt^2 + (H_1 H_2\ r^2 )^{1/5}(f^{-1} r^4 dr^2 + d\Omega_5^2)
\ee
where
\begin{eqnarray}
f &=& r^6 - \mu r^2  + \frac{1}{4L^2} H_1 H_2\\
H_i &=& r^4 + q_i\\
A^i_{(1)} &=& - q_i H_i^{-1} dt
\end{eqnarray}
and $d\Omega_5^2$ is the metric on a unit 5-sphere in flat space.
In the extremal case, $\mu=0$, there is a naked singularity at $r=0$. The lifted 11-dimensional supergravity solution inherits the naked singularity. We want to understand whether this singularity can be interpreted as a collection of giant gravitons. The results of the previous section will be used to probe the 11-dimensional supergravity solution with giant gravitons. We want to see whether the energy of a giant graviton probe is minimized at $r=0$, and whether it is a BPS minimum ($E_i = \tilde{P}_i/L$). We consider a giant graviton 2-brane probe which carries momentum in the $\phi_i$ direction, wraps the $\theta_{j\neq i}, \phi_{j\neq i}$ directions and is stationary in the extended directions, i.e.\ $\dot{x}^\mu =0$ except $\mu=0$. From Eq.~(\ref{Ei2}), the energy of this probe is given by
\be
E_i = \frac{1}{L} \sqrt{ - g_{00}}\ X_i \tilde{P}_i - \frac{1}{L} \tilde{P}_i A^i_0 = \frac{\tilde{P}_i}{L} \frac{f^{1/2} r + q_i}{H_i}
\ee
There are three distinct cases to consider:
\begin{enumerate}
\item{All $q_i=0$. BPS minimum at $r=0$ for probes coupling to both $A^{1}$ and $A^2$.}
\item{$q_1 \neq 0$, $q_2 =0$. Probe coupling to $A^1$ has a BPS minimum at $r=0$. Energy of probe coupling to $A^2$ diverges as $r\rightarrow 0$.}
\item{$q_1, q_2 \neq 0$. The energy of both probes saturates the BPS bound at $r=0$, but the gradient of the potential is non-zero at $r=0$.}
\end{enumerate}
Therefore, as in the previous case, it is only sensible to interpret the singly charged background as being sourced by giant gravitons. In this case a probe carrying the same type of charge as the background has a BPS minimum at $r=0$. In the doubly charged background this does not occur.

\section{Conclusions}
\label{conclusions}

We have demonstrated that massive charged particles in 4-dimensional
U(1)$^4$ gauged supergravity and 7-dimensional U(1)$^2$ gauged supergravity
correspond to giant gravitons in eleven dimensions. The existence of these
giant gravitons, which is a dynamical property of the brane interacting with
the background gauge field, does not depend on any specific solution of the
gauged supergravities or preservation of supersymmetry. So it appears that giant gravitons are closely related to the structure of the lift ansatze. These
calculations complement the results for 5-dimensional U(1)$^3$ gauged
supergravity obtained in \cite{Page:2002xz} as well as the results of
\cite{Das:2000ab} where it was first emphasised that giant gravitons can appear
in a variety of backgrounds, without requiring supersymmetry.

We also considered probing superstar geometries which are conjectured to be
sourced by a collection of giant gravitons. We obtained the same
qualitative results as in \cite{Page:2002xz} for the 5-dimensional case --
for the single charge solutions the probe calculation supported the conjecture
while for the multi-charges solutions the potential seen by the probes does not
have a BPS minimum
at the superstar singularity. Again the interpretation is not clear.
The results may indicate that the multi-charge solutions are not sourced by
giant gravitons, that the singularities of the multi-charge solutions are not
physical, or that the reduced supersymmetry of the multi-charge solutions
requires us to consider higher order curvature corrections in the probe
calculations. This is clearly an issue which requires further investigation.

Clearly giant gravitons are interesting objects within the context of gauged
supergravities and their lifts to 10- or 11-dimensional supergravity, as
well as in the context of gauge/gravity duality. In particular, we expect that
further investigation of these objects will lead to a better understanding of
consistent truncation to gauged supergravities. This may also illuminate the
role of solutions of gauged supergravities as opposed to solutions of 10- or
11-dimensional supergravity as the dual description of gauge theories.

\vskip.5in

\centerline{\bf Acknowledgements}
\medskip

We would like to thank David Page and Harald Svendsen for many helpful discussions. EJH is funded by the George Murray Scholarship (University of Adelaide) and the Overseas Research Students Awards Scheme.

\bibliographystyle{utphys}

\bibliography{references}

\providecommand{\href}[2]{#2}\begingroup\raggedright\begin{thebibliography}{10}

\bibitem{McGreevy:2000cw}
J.~McGreevy, L.~Susskind, and N.~Toumbas, ``Invasion of the giant gravitons
  from anti-de Sitter space,'' JHEP {\bf 06} (2000) 008,
\href{http://xxx.lanl.gov/abs/http://arXiv.org/abs/hep-th/0003075}{{\tt
  http://arXiv.org/abs/hep-th/0003075}}.
%%CITATION = HEP-TH 0003075;%%.

\bibitem{Balasubramanian:2001nh}
V.~Balasubramanian, M.~Berkooz, A.~Naqvi, and M.~J. Strassler, ``Giant
  gravitons in conformal field theory,'' JHEP {\bf 04} (2002) 034,
\href{http://xxx.lanl.gov/abs/http://arXiv.org/abs/hep-th/0107119}{{\tt
  http://arXiv.org/abs/hep-th/0107119}}.
%%CITATION = HEP-TH 0107119;%%.

\bibitem{Page:2002xz}
D.~C. Page and D.~J. Smith, ``Giant gravitons in non-supersymmetric
  backgrounds,'' JHEP {\bf 07} (2002) 028,
\href{http://xxx.lanl.gov/abs/http://arXiv.org/abs/hep-th/0204209}{{\tt
  http://arXiv.org/abs/hep-th/0204209}}.
%%CITATION = HEP-TH 0204209;%%.

\bibitem{Cvetic:1999xp}
M.~Cvetic {\em et.~al.}, ``Embedding AdS black holes in ten and eleven
  dimensions,'' Nucl. Phys. {\bf B558} (1999) 96--126,
\href{http://xxx.lanl.gov/abs/http://arXiv.org/abs/hep-th/9903214}{{\tt
  http://arXiv.org/abs/hep-th/9903214}}.
%%CITATION = HEP-TH 9903214;%%.

\bibitem{Duff:1999gh}
M.~J. Duff and J.~T. Liu, ``Anti-de Sitter black holes in gauged N = 8
  supergravity,'' Nucl. Phys. {\bf B554} (1999) 237--253,
\href{http://xxx.lanl.gov/abs/http://arXiv.org/abs/hep-th/9901149}{{\tt
  http://arXiv.org/abs/hep-th/9901149}}.
%%CITATION = HEP-TH 9901149;%%.

\bibitem{Sabra:1999ux}
W.~A. Sabra, ``Anti-de Sitter BPS black holes in N = 2 gauged supergravity,''
  Phys. Lett. {\bf B458} (1999) 36--42,
\href{http://xxx.lanl.gov/abs/http://arXiv.org/abs/hep-th/9903143}{{\tt
  http://arXiv.org/abs/hep-th/9903143}}.
%%CITATION = HEP-TH 9903143;%%.

\bibitem{Myers:2001aq}
R.~C. Myers and O.~Tafjord, ``Superstars and giant gravitons,'' JHEP {\bf 11}
  (2001) 009,
\href{http://xxx.lanl.gov/abs/http://arXiv.org/abs/hep-th/0109127}{{\tt
  http://arXiv.org/abs/hep-th/0109127}}.
%%CITATION = HEP-TH 0109127;%%.

\bibitem{Balasubramanian:2001dx}
V.~Balasubramanian and A.~Naqvi, ``Giant gravitons and a correspondence
  principle,'' Phys. Lett. {\bf B528} (2002) 111--120,
\href{http://xxx.lanl.gov/abs/http://arXiv.org/abs/hep-th/0111163}{{\tt
  http://arXiv.org/abs/hep-th/0111163}}.
%%CITATION = HEP-TH 0111163;%%.

\bibitem{Leblond:2001gn}
F.~Leblond, R.~C. Myers, and D.~C. Page, ``Superstars and giant gravitons in
  M-theory,'' JHEP {\bf 01} (2002) 026,
\href{http://xxx.lanl.gov/abs/http://arXiv.org/abs/hep-th/0111178}{{\tt
  http://arXiv.org/abs/hep-th/0111178}}.
%%CITATION = HEP-TH 0111178;%%.

\bibitem{Cvetic:1999ne}
M.~Cvetic and S.~S. Gubser, ``Phases of R-charged black holes, spinning branes
  and strongly coupled gauge theories,'' JHEP {\bf 04} (1999) 024,
\href{http://xxx.lanl.gov/abs/http://arXiv.org/abs/hep-th/9902195}{{\tt
  http://arXiv.org/abs/hep-th/9902195}}.
%%CITATION = HEP-TH 9902195;%%.

\bibitem{Liu:1999ai}
J.~T. Liu and R.~Minasian, ``Black holes and membranes in AdS(7),'' Phys. Lett.
  {\bf B457} (1999) 39--46,
\href{http://xxx.lanl.gov/abs/http://arXiv.org/abs/hep-th/9903269}{{\tt
  http://arXiv.org/abs/hep-th/9903269}}.
%%CITATION = HEP-TH 9903269;%%.

\bibitem{Das:2000ab}
S.~R. Das, S.~P. Trivedi, and S.~Vaidya, ``Magnetic moments of branes and giant
  gravitons,'' JHEP {\bf 10} (2000) 037,
\href{http://xxx.lanl.gov/abs/http://arXiv.org/abs/hep-th/0008203}{{\tt
  http://arXiv.org/abs/hep-th/0008203}}.
%%CITATION = HEP-TH 0008203;%%.

\end{thebibliography}\endgroup

\end{document}